\documentclass[superscriptaddress,nofootinbib,twocolumn]{revtex4}
\usepackage{amsmath,lscape,epsfig,ulem}
\usepackage{amsfonts}
\usepackage{amsmath}
\usepackage{amssymb}
\usepackage{graphicx}

\def\ii{\'{\i}}
\def\beq{\begin{equation}}
\def\eeq{\end{equation}}
\def\beqa{\begin{eqnarray}}
\def\eeqa{\end{eqnarray}}
\def\ban{\begin{eqnarray*}}
\def\ean{\end{eqnarray*}}
\def\bi{\begin{itemize}}
\def\ei{\end{itemize}}

\begin{document}

\title{ The QCD Critical End Point Under Strong Magnetic Fields}

\author{Sidney S. Avancini}
\affiliation{Depto de F\'{\i}sica - CFM - Universidade Federal de Santa
Catarina  Florian\'opolis - SC - CP. 476 - CEP 88.040 - 900 - Brazil}
\author{D\'{e}bora P. Menezes}
\affiliation{Depto de F\'{\i}sica - CFM - Universidade Federal de Santa
Catarina  Florian\'opolis - SC - CP. 476 - CEP 88.040 - 900 - Brazil}
\author{Marcus B. Pinto} \email{marcus@fsc.ufsc.br}
\affiliation{Depto de F\'{\i}sica - CFM - Universidade Federal de Santa
Catarina  Florian\'opolis - SC - CP. 476 - CEP 88.040 - 900 - Brazil}
\author{Constan\c ca Provid\^encia}
\affiliation{Centro de F\ii sica Computacional - Department of Physics -
University of Coimbra - P-3004 - 516 - Coimbra - Portugal}

\begin{abstract}
We use the three-flavor Nambu--Jona-Lasinio model,{ which includes strangeness and quark physical masses} in the mean field approximation,  to investigate the influence of strong magnetic fields on the QCD phase diagram covering the whole $T-\mu$ plane.  It is found that the size of the first 
order transition line increases as the field strength increases so that a larger coexistence region for hadronic and quark matter should be expected for strong magnetic fields. The location of the critical end point is also  affected by the presence of magnetic fields which invariably increase the temperature value at which the first order line terminates. On the other hand, the critical end point chemical potential value displays a subtle   oscillation   around the $B=0$ value for magnetic fields within the $10^{17}-10^{20} \, {\rm G}$ range. These findings may have non trivial consequences for the physics of magnetars and heavy ion collisions.

\end{abstract}

\maketitle

\vspace{0.50cm}
PACS number(s): {24.10.Jv,11.10.-z, 25.75.Nq}
\vspace{0.50cm}

Magnetic catalysis is an interesting phenomenon with direct consequences on the QCD chiral symmetry breaking (CSB) mechanism. At vanishing $T$ and $\mu$  one observes, within effective quark models, that the presence of a magnetic field ($B$) stabilizes the chirally asymmetric vacuum by antialigning the helicities of a quark--anti-quark pair (${\bar q} q$)  which is then bound by the strong interaction. This contrasts with the behavior of the BCS ground state of an ordinary superconductor where $B$ favors the alignment of the spins for an electron pair opposing pair formation. As a consequence of magnetic catalysis, the quark condensate, and hence the  quark effective mass, assumes higher values as $B$ increases. Predictions based on the Nambu--Jona-Lasinio model (NJL) \cite{njl} suggest that the  effective quark mass raises by approximately  $30$ to $40 \%$ when $B$ increases from $10^{17}$ to $ 3 \times 10^{19}$ Gauss, i.e., by about two orders of magnitude \cite{prc,hotnjl}. It is important to stress that the effective masses at vanishing $B$ coincide with the ones obtained for $B=10^{17}$ G or lower. 
The fact that magnetic fields induce  magnetic catalysis, enhancing  CSB, naturally leads to the question of how these fields would influence chiral phase transitions at finite temperatures and/or chemical potentials. 
Of course this is more than merely an academic question since one can immediately point out at least two realistic scenarios whose physics  are influenced by the behavior of  strongly interacting matter under intense magnetic fields. The first refers to the high temperature  and low chemical potential  regime relevant to non central heavy ion collisions where huge fields, of the order $|eB| \ge m_\pi^2 \sim 10^{18} \, {\rm G}$,  can be created by heavy ion currents due to the spectator nucleons. The field intensity depends on the centrality and beam momentum so that  $|eB| \approx 5 m_\pi^2$ can be reached at RHIC while $|eB| \approx 15 m_\pi^2$ can be reached at the LHC. Although time dependent and short-lived \cite{kharzeev_09} these fields may have a non negligible effect on the transition \cite{lattice}. 
Strong magnetic fields in heavy ion collisions at beam energies ranging from 0.2 to 2 GeV/nucleon, which cover
most of the accelerators in the world,  were considered in \cite{bao-an2011}, where the authors conclude that  particles with low masses and their differential ratios (as $\pi^-/\pi^+$, for example) can be used as possible probes of the density dependence of the symmetry energy, another topic of interest in the recent literature.  
The  second scenario refers to the cold baryon-dense matter which forms compact stellar objects such as magnetars. While ordinary neutron stars bear a magnetic field of the order of $10^{9}-10^{12} \, {\rm G}$, magnetars, believed to be the source of intense gamma and X rays, bear fields  of the order of $10^{13}-10^{15} \, {\rm G}$
at their surface reaching values up to 3 to 5 times of magnitude greater in their center \cite{magnetars}.
As far as phase transitions are concerned, these two physically appealing
situations are located at the high-$T$--low-$\mu$ (relativistic heavy ion
collisions) and low-$T$--high-$\mu$ (neutron stars) extremes of the QCD phase
diagram while the intermediate region lies within the capabilities of 
experiments such as the low energy scan in HIC at RHIC, FAIR at GSI, NICA at
JINR, and J-PARC at JAERI where, eventually, high magnetic fields will also be
reached. 

The determination of the QCD phase diagram, even at vanishing $B$, is still a matter of great theoretical and experimental activities. In this case, powerful lattice simulations have established that at vanishing baryon densities there is no true phase transition from hadronic matter to a quark gluon-plasma but rather a very rapid raise in the energy density signaling a crossover characterized by a pseudocritical temperature, $T_{\rm pc}$, which is expected to be within the  $150-200 \, {\rm MeV}$ range, with systematic errors \cite{lattice}. The situation is less clear for the finite chemical potential region since, so far, there is no reliable information avaliable from lattice QCD  evaluations. Nevertheless, most finite $\mu$ lattice extrapolations for the $\mu=0$ Columbia plot indicate that the critical first order surface (on the $m_{u,d}-m_s-\mu$ plane) will hit the physical current mass values at some finite $\mu$ thereby characterizing a critical end point (CEP) \cite {fukuhatsuda}.

Therefore, the use of effective models for QCD  constitutes the pragmatic approach to treat the finite chemical potential region, especially at very low temperatures. Within this spirit, the three-flavor version of the NJL has been widely used in investigations which aim to analyze the phase structure of chiral transitions. Using  a standard parametrization, at $B=0$, this model predicts a phase diagram \cite {pedro2008} which  is in accordance with what is the most general current belief. Namely, the model predicts that at vanishing chemical potential a crossover takes place with a $T_{\rm pc}$ which is in good agreement with the lattice results. At zero temperature, a first order transition occurs for $\mu \sim M_B/3$, where $M_B$ is the baryon mass,  so that a first order transition line emerges at smaller $\mu$ as $T$ departs from zero. Since at $\mu \cong 0$ a crossover is predicted, this line cannot cover the whole $T-\mu$ plane and will terminate at a CEP  located  at intermediate values of $T$ and $\mu$. Considering this picture as the $B=0$ benchmark one can further analyze how a magnetic field will influence the expected QCD chiral transition. 
So far, most model applications carried out with this aim were performed at $\mu=0$  with  the two-flavor linear $\sigma$ model (LSM) \cite {eduardo} as well as with the two-flavor NJL model \cite {ruggieri} predicting that the crossover takes place at higher $T_{\rm pc}$ when $B \ne 0$. Another interesting outcome from these investigations shows that  the chiral and deconfining lines can split at finite $B$ when the models also include the Polyakov loop. On the lattice side, a two flavor simulation, which used quark current masses corresponding to pion masses on the range 200-400 MeV has shown that the deconfinement and chiral symmetry raise by just a few percent even for very high fields and have not observed any splitting \cite {1stlattice}. 
However, very recently another lattice simulation has considered 2+1 flavors and physical values for the quark current masses  predicting that the pseudo critical temperature should decrease with increasing $B$ \cite{lattice}. This also constitutes an interesting result if one recalls that the running of the strong coupling in the presence of $B$ may turn the magnetic catalysis effect around, as suggested in Ref. \cite {miransky}.
 
Despite  the progress made at vanishing baryon density very little has been
done to determine the influence of magnetic fields on the whole $T-\mu$
plane. One such investigation was carried out in Ref. \cite {andersen} using
the standard two flavor LSM but when the vacuum is  included, as it should be \cite {eduardo}, the whole $T-\mu$ plane is dominated by the crossover just as it happens at $B=0$. On the contrary, as shown in \cite{inagaki}, the standard two flavor version of the NJL model in the chiral limit does not share this property allowing for the observation of how a magnetic field influences the location of the tricritical point. In Ref. \cite {inagaki} a Fock-Schwinger proper time method was used in the chiral limit. 
The authors have observed a curious pattern between magnetic catalysis and the Haas-van Alphen effect, which leads
to oscillations in the first order transition line.
The complexity of the phase diagram obtained with several tricritical
points is probably due to the zero mass quark limit and further
investigation with realistic current quark masses is needed. As emphasized in
\cite{muller}, strange quarks play a crucial role in shaping the phase diagram
of QCD since their mass, $m_s$, controls the nature of the chiral and deconfinement transition. 
At the same time, this mass has also an important effect on the stability limit of neutron stars and on the possible existence of a quark core in collapsed stars. For these objects, the establishment of the critical first order line is of utmost importance once the possibility for  hybrid and quark star formation depends on the quark matter nucleation process. The determination of the first order line, at $B\ne0$,  may also be relevant in the framework of studies related to thermal nucleation of quark droplets in hadronic matter found in protoneutron stars whose temperatures are of the order 10--20 MeV. 

In this Letter we report on the main results obtained by scanning  the whole $T-\mu$ plane paying special attention  to the CEP and to the first order coexistence region. We do not consider the Polyakov loop since its effects are more pronounced at the high temperature and low chemical potential region \cite {eduardo,ruggieri}.
In \cite{pedro2009} it was shown that, although it changes considerably the values of the temperature and chemical potential related to the CEP, the qualitative behavior of the phase transition was not modified. 

 By mapping the first order line on the $T-\mu$ plane we can automatically predict the location of the CEP for different values of $B$.  At vanishing $\mu$ we reproduce the results known from the other model applications (like those in Refs. \cite {eduardo,ruggieri}) predicting that the $B=0$ crossover will take place at a higher $T_{\rm pc}$
for $B \ne 0$. The new results reported here emerge at the finite baryonic density domain where we observe that the presence of a magnetic field favors the presence of latent heat at higher temperatures, invariably increasing the size of the first order transition region so that the CEP temperature, $T_{\rm CEP}(B)$, increases with $B$.
We also show that the critical chemical potential,  $\mu_{\rm crit}(B)$, which determines the low temperature first order transition displays a non trivial behavior moving around  the $B=0$ value for magnetic fields in the range $10^{17} - 10^{20}$ G.
To the best of our knowledge this is the first time such predictions are made within a more realistic approach, which takes strangeness
and physical masses into account.
 
\begin{figure*}[htb]
\begin{center}
\begin{tabular}{cc}
\includegraphics[width=0.45\linewidth]{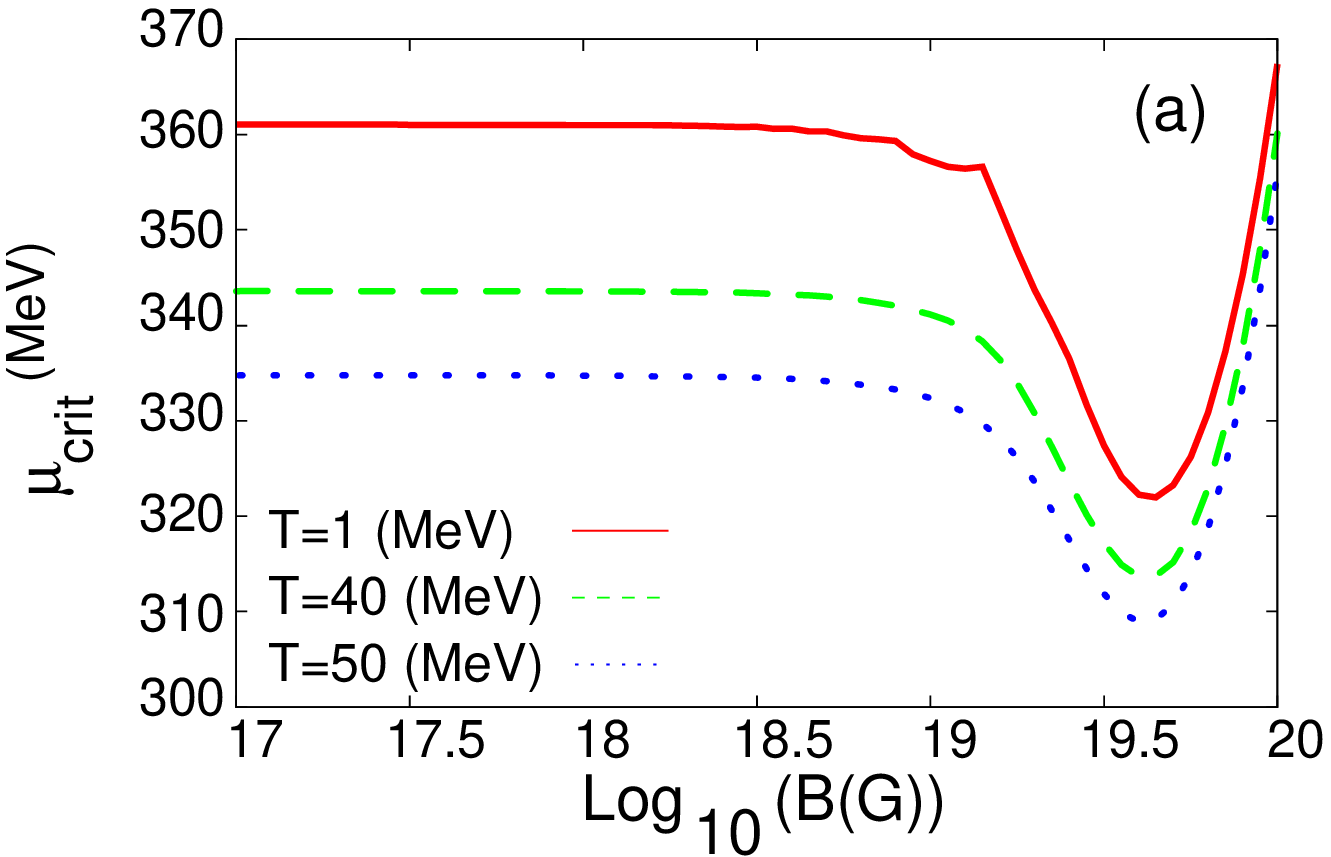}  &
\includegraphics[width=0.4\linewidth]{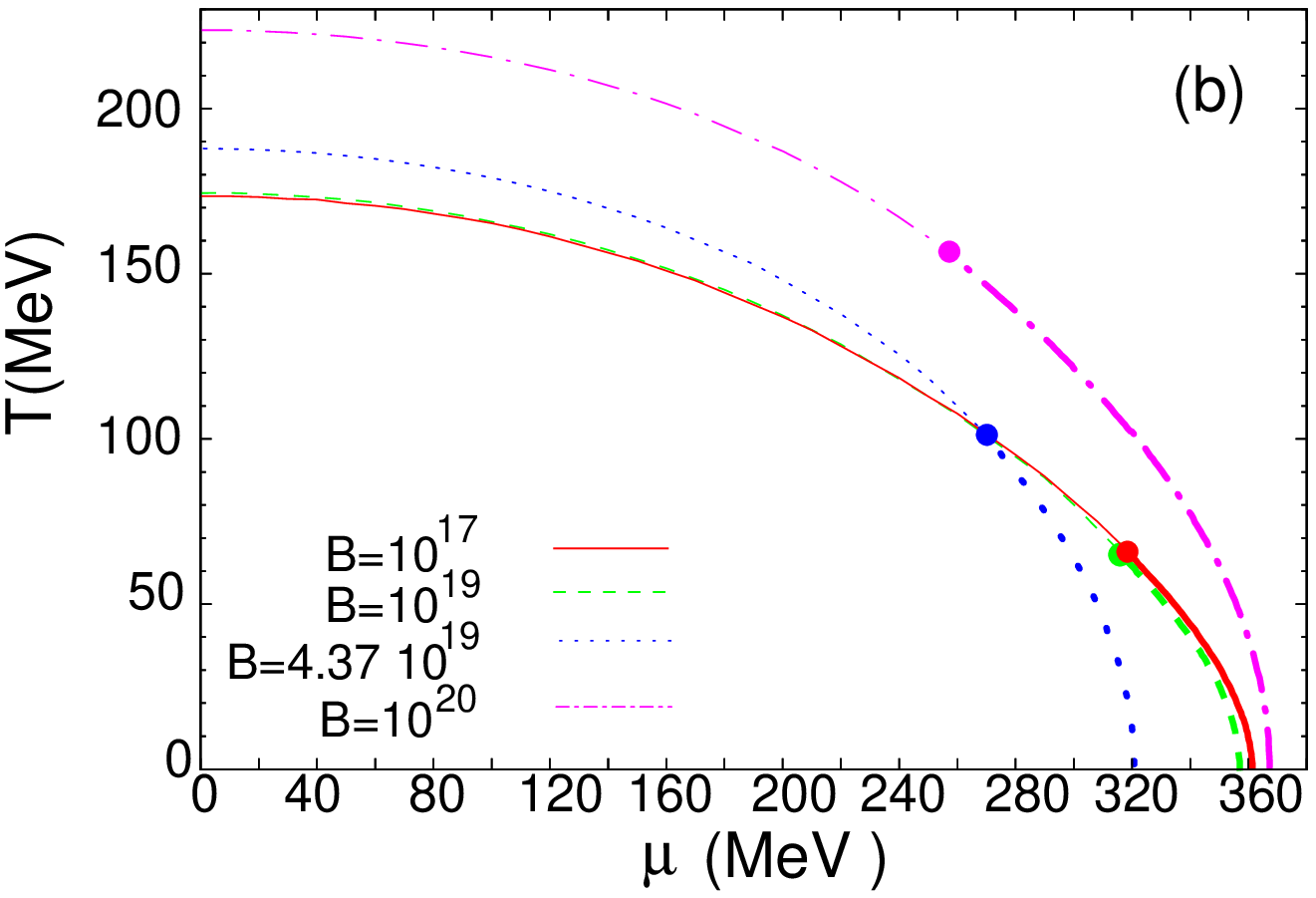} \\
\end{tabular} 
\end{center}
\caption{ a) Critical chemical potential in terms of the magnetic field
  for different values of the temperature; b) Phase diagram on the $T-\mu$ plane.
The thick curves represent first order transition lines which terminate at a
critical end point identified with a full dot  and
the thin lines represent a crossover.}
\label{fig1}
\end{figure*}

In order to consider (three flavor) quark matter subject to 
strong magnetic fields we introduce the following Lagrangian density
\begin{equation}
{\cal L} = {\cal L}_{f} - \frac {1}{4}F_{\mu \nu}F^{\mu \nu},
\end{equation}
where the quark sector is described by the  su(3) version of the
Nambu--Jona-Lasinio model which includes scalar-pseudoscalar
and the t'Hooft six fermion interaction that
models the axial $U(1)_A$ symmetry breaking  \cite{njlsu3}:
\begin{equation}
{\cal L}_f = {\bar{\psi}}_f \left[\gamma_\mu\left(i\partial^{\mu}
- q_f A^{\mu} \right)-
{\hat m}_c \right ] \psi_f ~+~ {\cal L}_{sym}~+~{\cal L}_{det}~,
\label{njl}
\end{equation}
with ${\cal L}_{sym}$ and ${\cal L}_{det}$  given by \cite{buballa}:
\begin{equation}
{\cal L}_{sym}~=~ G \sum_{a=0}^8 \left [({\bar \psi}_f \lambda_ a \psi_f)^2 + ({\bar \psi}_f i\gamma_5 \lambda_a
 \psi_f)^2 \right ]  ~,
\label{lsym}
\end{equation}
\begin{equation}
{\cal L}_{det}~=~-K \left \{ {\rm det}_f \left [ {\bar \psi}_f(1+\gamma_5) \psi_f \right] + 
 {\rm det}_f \left [ {\bar \psi}_f(1-\gamma_5) \psi_f \right] \right \} ~,
\label{ldet}
\end{equation}
where $\psi_f = (u,d,s)^T$ represents a quark field with three flavors, ${\hat m}_c= {\rm diag}_f (m_u,m_d,m_s)$ is the corresponding (current) mass matrix while $q_f$
represents the quark electric charge,{  $\lambda_0=\sqrt{2/3}I$  where
$I$ is the unit matrice in the three flavor space, and
$0<\lambda_a\le 8$ denote the Gell-Mann matrices.
The model is not renormalizable, and as a regularization scheme for  the
divergent ultra-violet integrals  we use a sharp cut-off, $\Lambda$, in
3-momentum space. The parameters of the model, $\Lambda$, the coupling 
constants $G$ and $K$
and the current quark masses $m_u^0$ and $m_s^0$ are determined  by fitting
$f_\pi$, $m_\pi$ , $m_K$ and $m_{\eta'}$ to their empirical
values. Throughout this paper we consider 
$\Lambda = 631.4 \, {\rm MeV}$ , $m_u= m_d=\,  5.5 \,{\rm MeV}$,
$m_s=\,  135.7\, {\rm MeV}$, $G \Lambda^2=1.835$ and $K \Lambda^5=9.29$
as in \cite{hatsuda}.

As  usual, $A_\mu$ and $F_{\mu \nu }=\partial
_{\mu }A_{\nu }-\partial _{\nu }A_{\mu }$ are used to account
for the external magnetic field. We consider a  static and 
constant magnetic field in the $z$ direction, $A_\mu=\delta_{\mu 2} x_1 B$.

The thermodynamical potential for the three flavor quark sector,
$\Omega_f$ is written as
\begin{eqnarray}
\Omega_f(T,\, \mu,\, B)&=& 2G \sum_{i=u,\,d,\,s} \phi_i^2 -4K \, \phi_i \, \phi_j \, \phi_k\nonumber\\
&+&\left(\Omega_f^{vac}+\Omega_f^{mag}+\Omega_f^{med}\right),
\end{eqnarray}
where the vacuum ($\Omega_f^{vac}$), the magnetic ($\Omega_f^{mag}$), the medium contributions
($\Omega_f^{med}$) and the condensates $\phi_i$ have been evaluated with great detail in Ref. \cite{prc,hotnjl}. 
The effective quark masses are solutions of  the gap equations obtained from 
the minimization of the thermodynamic potential with respect to the effective masses, 
\begin{equation}
 M_i=m_i - 4 G \phi_i + 2K \phi_j \phi_k, 
 \label{mas}
\end{equation}
with $(i,j,k)$ being any permutation of $(u,d,s)$. In this work we consider $\mu=\mu_u=\mu_d=\mu_s$.

The first order phase transition line is obtained by enforcing the  Gibbs
conditions while the cross over region is determined by the maximum of $-d\phi_m/dT$ ($m=u,d$). 

\begin{figure*}[htb]
\begin{center}
\includegraphics[width=0.8\linewidth]{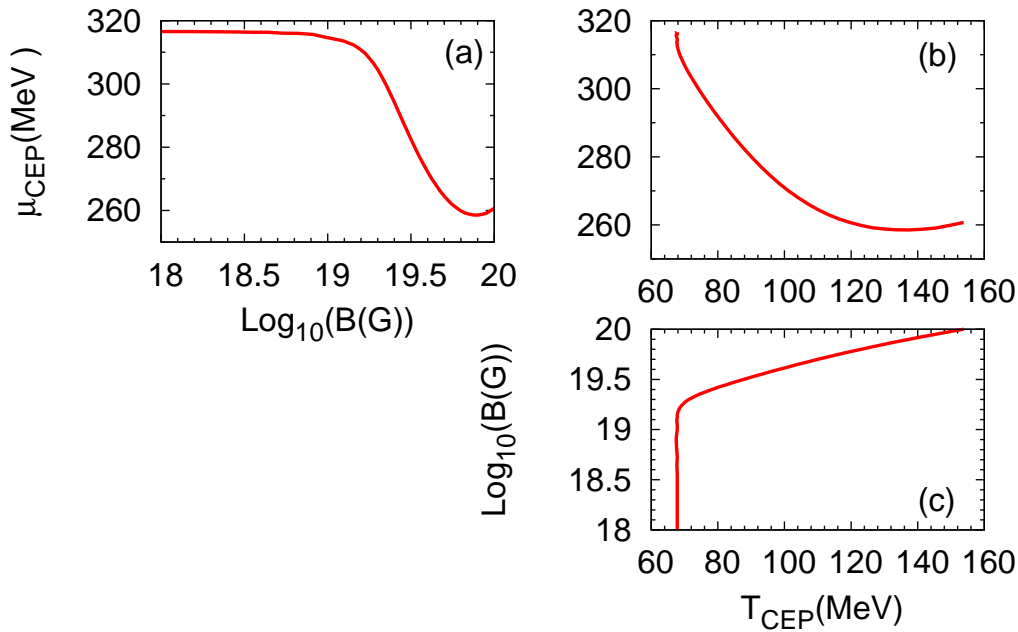} 
\end{center}
\caption{Critical end point chemical potential $\mu_{CEP}$ in terms of a) the
  magnetic field, and b) the critical end point temperature $T_{CEP}$;  
c) $T_{CEP}$ as a function of $B$.}
\label{fig2}
\end{figure*}

In Fig. \ref{fig1}a we plot the critical chemical potential as a function
of the magnetic field for different low temperature values where the first order transitions take place. For a fixed magnetic
field, the critical chemical potential ($\mu_{\rm crit}$) is approximately constant up to magnetic
fields of the order of $10^{19}$ G, where it reduces drastically
reaching a minimum value around $3\times 10^{19}$ G and then it
increases again.  This occurs precisely when just one Landau level
is filled. One can clearly see the effects of the filling of the Landau levels
at low temperatures, where the van Alphen oscillations are seen. 

The effects of the behavior observed in Fig. \ref{fig1}a are reflected in
the curves shown in Fig. \ref{fig1}b, where a $T-\mu$ phase diagram is shown for different values of $B$. 
As the magnetic field increases, 
the line of the first order phase transition also increases. The line moves 
towards a lower chemical potential and then back to higher chemical potentials for increasing values of $B$.
The highest point in each first order transition curve is the CEP, indicated with a full dot.
 The curve obtained for $B=10^{17}$ 
G coincides exactly with the results presented in \cite{pedro2008}  for
a zero magnetic field. 
The pattern observed in Fig. \ref{fig1}b can  be seen with more detail in Fig.
 \ref{fig2} where the dependence of $\mu_{CEP}$ and $T_{CEP}$, which characterize the CEP,
are shown as a function of the magnetic field $B$. Fig. \ref{fig2}a shows that $\mu_{CEP}$ essentially decreases with $B$, with
a lower decrease for fields stronger than $10^{19}$ G. Just below $10^{20}$ G, $\mu_{CEP}$ starts to
increase. Fig. \ref{fig2}b represents $\mu_{CEP}$ versus $T_{CEP}$, each point corresponding to a different value of the magnetic field:
the smallest value of the critical chemical potential, 258.5 MeV, corresponding to a $T_{CEP} \simeq 137$ MeV occurs for
$B=8\times 10^{19}$. On the other hand, $T_{CEP}$ always increases with the increase of $B$, mainly above $10^{19}$ G.

In summary, our results show that, at $\mu=0$, $T_{\rm pc}$ increases with $B$ in agreement with most model calculations  \cite {eduardo, ruggieri} and an early lattice simulation \cite{1stlattice} which is not surprising since we are dealing with the NJL within the MFA. On the other hand, the main contribution of our work concerns the critical end point and the first order coexistence region for which magnetic effects become noticeable for fields greater than $10^{17} \, {\rm G}$. In this case, one observes an increase on the size of the first order transition line with $T_{\rm CEP}$ moving to higher values as $B$ increases. This result can be of interest for the physics of heavy ion collisions at intermediate energies. We have also observed that, due to the van Alphen effect, the first order transition lines display a non trivial behavior oscillating around the $B=0$ line. This implies that, as $B$ increases, criticality will be achieved at chemical potential values which can be greater or smaller than the $B=0$ value, $\mu_{\rm crit}  \simeq 360 \, {\rm MeV}$. 
These findings may have direct consequences regarding the dynamics of the first order phase transition which influences the quark matter nucleation process. Although we have attempted to perform a realistic application, by considering three quark flavor and physical mass values, it is important to recall that at temperatures higher than 100 MeV the effects of the Polyakov loop
may be important \cite{celia}. The Polyakov loop field is going to be more carefully 
considered in a forthcoming paper but we belive that these seminal qualitative results
are not going to be altered.

{\bf Ackowledgements}: This work was partially supported by the Capes/FCT n. 232/09 bilateral 
collaboration, by CNPq and FAPESC (Brazil), by FCT and FEDER (Portugal) 
under the project PTDC/FIS/113292/2009 and  by Compstar, an ESF Research
Networking Programme.

\end{document}